# Routing of *Physarum polycephalum* "signals" using simple chemicals

Ben de Lacy Costello[1,2] and Andrew Adamatzky[2]

[1]Institute of Biosensing Technology; University of the West of England; Bristol, UK
[2]Unconventional Computing Group; University of the West of England; Bristol, UK

In previous work the chemotaxis towards simple organic chemicals was assessed. We utilise the knowledge gained from these chemotactic assays to route *Physarum polycephalum* signals at a series of junctions. By applying chemical inputs at a simple T-junction we were able to reproducibly control the path taken by the plasmodium of *P. Polycephalum*. Where the chemoattractant farnesene was used at one input a routed signal could be reproducibly generated i.e. *P. Polycephalum* moves towards the source of chemoattractant. Where the chemoattractant was applied at both inputs the signal was reproducibly split i.e. at the junction the plasmodium splits and moves towards both sources of chemoattractant. If a chemorepellent was used then the signal was reproducibly suppressed i.e. *P. Polycephalum* did not reach either output and was confined to the input channel. This was regardless of whether a chemoattractant was used in combination with the chemorepellent showing a hierarchy of inhibition over attraction. If no chemical input was used in the simple circuit then a random signal was generated, whereby *P. Polycephalum* would move towards one output at the junction, but the direction was randomly selected.

We extended this study to a more complex series of T-junctions to explore further the potential of routing *P. Polycephalum*. Although many of the "circuits" were completed effectively, any errors from the implementation of the simple T-junction were magnified. There were also issues with cascading effects through multiple junctions. For example signal splitting could be reproducibly initiated at the first junction but not at subsequent junctions. Also to implement more complex circuits of this type within a given area junction size is reduced and thus the distance between chemical sources is decreased meaning that the routing of the signal is more challenging. This work highlights the potential for exploiting chemotaxis to achieve complex and reliable routing of *P. Polycephalum* signals. This may be useful in implementing computing algorithms, design of autonomous robots and directed material synthesis.

In additional experiments we showed that the application of chemoattractant compounds at specific locations on a homogeneous substrate could be used to reliably control the spatial configuration of *P. Polycephalum*. This may have applications in implementing geometric calculations and in robot navigation tasks such as mapping chemical plumes.

## Introduction

*Physarum polycephalum* is a true acellular slime mold that belongs to the species of order *Physarales*, subclass *Myxo-gastromycetidae*, class *Myxomycetes*, division *Myxostelida*. The life cycle of *P. polycephalum* possesses a plasmodial phase where it exists as a single cell with a large number of diploid nuclei. The plasmodium is yellow coloured and moves like a giant amoeba, deploying a network of protoplasmic tubes whilst searching for food, which typically consist of bacteria, spores and micro-particles.[1] Any fragment of a plasmodium restores the integrity of the surrounding membrane and resumes the contractile and locomotive activities, therefore, fragments of standard size and shape can be used in chemotactic assays.[2,3]

Cytoplasm is streamed rhythmically back and forth through a network of tubular elements, circulating nutrients and chemical signals and forming pseudopods that allow the organism to navigate around and respond to its environment. The plasmodium propagates according to the position of nutrients but also in response to external gradients in light level and humidity. *P. polycephalum* will also propagate according to gradients in certain chemical species, either chemoattractants or chemorepellents. The *P. polycephalum* plasmodium is a model system for studying non-muscular motility, and its chemotactic behavior has been well documented.[4-8] In particular, substances causing negative taxis (chemorepellents) were shown to increase the period of contractility and to decrease the area of spreading when present uniformly within the substrate.[6,7]

Experimental studies confirmed that the following substances acted as chemoattractants for the plasmodium, glucose, galactose, maltose and mannose,[8,9] peptones,[8,10] the amino acids phenylalanine, leucine, serine, asparagine, glycine, alanine, aspartate, glutamate; and threonine,[11-13] phosphates, pyrophosphates, ATP and c AMP and thorium nitrate.[14] A plasmodium is allegedly indifferent to fructose and ribose.[8,9] Whereas, the following compounds have been found to act as chemorepellent molecules, sucrose and inorganic salts such as the chloride salts of (K, Na, $NH_4$, Ca, Mg, La)[14,15] and tryptophan.[13] Therefore, it is clear that the nutritional value of the substance is not paramount in determining either chemoattractant or chemorepellent properties.[12] Although recently there has been renewed interest in the question of nutritional value and chemotaxis.[16] For some substances, the effect on the plasmodium can be determined by the proximity of the organism to the source (or the concentration of the source), meaning that some substances can act as both chemoattractant and chemorepellent molecules. An example is the sugars galactose and mannose, which are reported to act as chemoattractants[8,9] and chemorepellents that inhibit motion.[17]

Recently it was found[18] that the plasmodium is strongly attracted to herbal medicines. Laboratory, experiments were undertaken on the plasmodium's binary choice between samples of dried herbs/roots: *Valeriana officinalis*, *Humulus lupulus*, *Passiflora incarnate*, *Lactuca virosa*, *Gentiana lutea* and *Verbena officinalis*. A hierarchy of chemo-attractive force was calculated from the binary interactions and it was found that *Valeriana officinalis* was the strongest chemo-attractant for *P. polycephalum* of the substances tested. However, it is unclear which component is causing the chemo-attractive effect, although actinidine a component of valerian root is known to have a chemo-attractive effect on a number of animal species[19]

In previous work[2] we found that *P. polycephalum* exhibited positive and negative chemotaxis to a range of simple volatile organic compounds. The chemoattractive compounds in order of strength of action were as follows: Farnesene > β-myrcene > tridecane > limonene > p-cymene > 3-octanone > β-pinene > m-cresol > benzylacetate > cis-3-

hexenylacetate. The chemorepellent compounds in order of strength of action were: nonanal > benzaldehyde > methylbenzoate > linalool > methyl-p-benzoquinone > eugenol > benzyl alcohol > geraniol > 2-phenylethanol. In further work[20] it was shown that these chemicals cause a distinct and measureable shift in the frequency and amplitude of the plasmodium generated oscillations in electrical potential. Furthermore, that this change is dependent on a chemoattractive, chemorepellent or neutral effect.

Recently, the plasmodial phase of *P. polycephalum* has been used extensively as a biological computing substrate. It has been used to solve a wide range of computationally hard problems such as maze-solving, the travelling salesman problem, calculation of optimal graphs, construction of logical gates and arithmetic circuits, sub-division of spatial configurations of data points and robot control.[21-30]

Routing of signals plays a critical role in the design of modern electronic circuits and computational chips[31-32]. Indeed the routing of signals is a major barrier to making smaller, faster and energy efficient chips and/or the integration of chips into computing architectures. It is important to be able to precisely manipulate data signals within a computing chip in order to preserve data integrity. Directivity and timing are important factors in signal routing. It is also important to be able to split, fuse, bend and filter data signals. It is beneficial to have on-chip integrated methods for routing signals. There is also an increasing trend to adopt biologically inspired methodologies to tackle problems of routing within conventional and unconventional computing approaches.[33-36]

In previous work the routing of *P. polycephalum* was studied using a flow of electrical current.[37] Also in recent related work the routing of plant roots through a Y junction was investigated using the volatile chemicals diethyl ether, ethephon and methyl jasmonate[38].

This paper details experiments which aim to use the acquired knowledge concerning the chemotactic effect of simple organic chemicals to control the movement of *P. polycephalum* through a series of junctions. This routing of signals could be useful in designing computing circuits modulated by chemicals and other external stimuli. It may also be useful in designing motion control circuits for robots particularly where taxis towards a target analyte is desired.

**Experimental**

**Culturing of *Physarum polycephalum*.** The true slime mold, the plasmodium of *Physarum polycephalum* (strain HU554 × HU560), was cultured with oat flakes on a 1% agar gel at 25°C in the dark. To obtain large quantities of inoculated oat flakes for chemotactic assays the plasmodial phase of *P. polycephalum* was cultivated in large plastic containers on filter paper that was wetted with 5 mls of de-ionized water. Any excess water was removed from the container. A source of food was added in the form of 50 g of rolled oats per container (Organic rolled oats). These containers were covered in order to retain moisture and kept in the dark at 25°C until required. These cultures were checked daily and water added if required. Sub-cultures were taken every 2–3 days to establish consistent cultures for ongoing experiments. Sub-culturing simply involved the removal of colonized oat flakes from the main culture and addition to 9 cm petri dishes containing non-nutrient agar. The colonised oat flake was placed in the centre of the petri dish and then additional uncolonised oat flakes were added around the periphery of the dish separated by a few cm. These dishes were then sealed and stored in the dark. The main cultures were also renewed regularly from the sub-cultures.

**Routing of Physarum signals.** Experiments were performed in 9 cm diameter polystyrene Petri dishes. A 1% solution of agar (Select Agar, Sigma) was added to each Petri dish to give a depth of approximately 2 mm. Three different experimental approaches to investigate the routing of *P. polycephalum* signals were undertaken.

**Simple T-shaped junction**

In this approach T-shaped junctions (see Figure 1.) were formed by cutting a T-shape with 1cm wide channels from a single sheet of agar within a petri dish. The length of each input segment was *circa* 4cm. An oat flake colonised by *P. polycephalum* was placed at the bottom of the vertical junction. The petri dish was then sealed and left in the dark for 2 hours prior to the addition of any chemicals. After two hours Squares of filter paper (circa 1.0 cm$^2$) were cut and placed at the ends of the horizontal junctions. Then 50µl of certain volatile organic chemicals (VOCs) which had been shown previously to have a chemotactic effect on *P. polycephalum* were added to the filter paper. These chemicals included α-farnesene (Sigma Aldrich UK) which was shown to have a relatively strong positive chemotactic response, cis-3-hexenyl acetate (Sigma Aldrich UK) which was shown to have a weak negative chemotactic effect. Alternatively the filter paper was left without any chemical. All possible binary combinations of the three reagents were implemented. The analysis was repeated 10 times to assess the reproducibility of the observed effects.

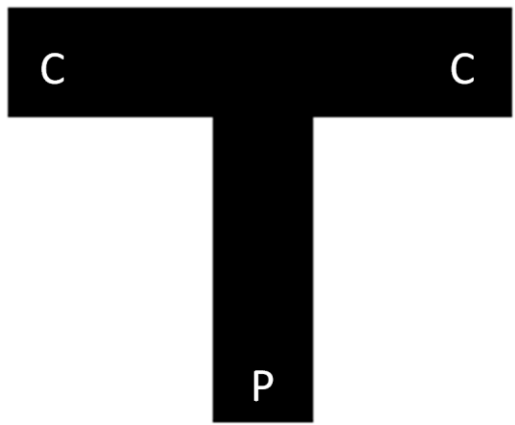

**Figure 1.** T-shaped junction used in *P. polycephalum* routing experiments. P marks the point of *P polycephalum* inoculation, C marks the points of chemical input and the target outputs for *P. Polycephalum*.

**Compound T-shaped junction**

The analysis described above was extended to include a more complex junction with a series of decision points. This junction consisted of the original T-shaped junction, but at the terminus of each horizontal junction there was a replicate T-shaped junction. Therefore, rather than two chemical inputs there were now six possible inputs. To accommodate this complex junction within a single petri dish it was necessary to reduce the width of the channels to 0.8 cm, and the length of the channels to 1.6cm. The experiment was undertaken using the method described above re the inoculation and incubation of *P. polycephalum* prior to addition of the chemicals. In this analysis an inhibitor was not used and therefore, all possible combinations of farnesene and the input of no chemical were explored. Again the analysis was repeated 10 times.

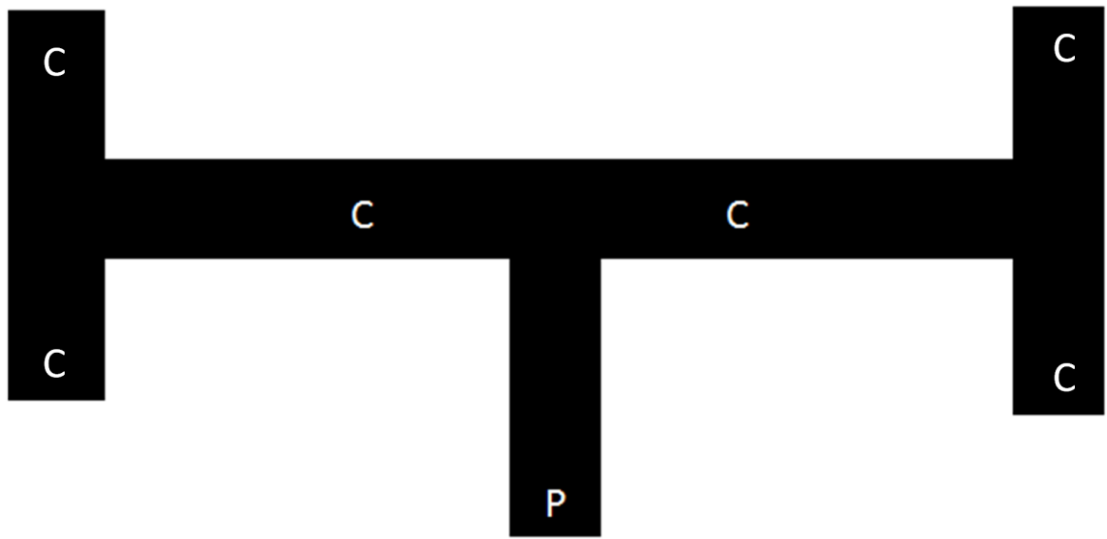

**Figure 2.** Compound T-junction used in P. Polycephalum routing experiments. P marks the point of *P polycephalum* inoculation, C marks the points of chemical input and the target outputs for *P. Polycephalum*.

**Control of spatial arrangement of Physarum cultures**

This experiment did not use junctions cut from agar, but simply used a single petri dish containing a thin layer of agar. In this case Physarum was inoculated on an oat flake in the centre of the petri dish and left to incubate for 2 hours. Four 1cm$^2$ filter paper squares were placed at the the NSWE compass points within the petri dish. The addition of farnesene or no chemical to these squares using all possible combinations was investigated.

**Analysis of results**

In all experiments the Petri dishes were sealed using parafilm and kept in the dark. They were checked at six hourly intervals for evidence of any effect. Usually between 24–48 h, the behavior was established and recorded. All dishes were scanned in batches of six using a flatbed scanner (HP Scanjet 5590) interfaced to a PC.

**Results and Discussion**

**Simple T-shaped junction**

It was found that by using a combination of chemoattractant (activator A, farnesene) and chemorepellent (inhibitor I, cis-3-hexenyl acetate) chemicals in combination with a zero input which was the addition of no chemical (N, neutral) that it was possible to reproducibly control the routing of *P. polycephalum* signals through a simple T-shaped junction. **Figure 3**. Shows selected results from the experiment. It shows that an inhibitor chemical can be used to suppress the *P. polycephalum* signal so that no output is obtained in either output channels of the T-junction. This effect was observed for all the following possible inputs (AI, IA, II, IN and NI) with the output being (00). As shown in Figure 3. the level of suppression is different, depending on the other input in combination with the inhibitor. Thus in the top left hand image the *P. polycephalum* signal is confined to the input channel but has advanced a significant distance from the inoculation site. This is due to the combination of activator with inhibitor. Compare this to the combination of inhibitor with no chemical input were the signal is confined to the locality of the inoculation site. Where inhibitor was placed at both inputs the signal did not propagate from the inoculation site. These observations could be exploited further in the design of more complex junctions/circuits. For example the *P. polycephalum* signal is only confined to the input channel where the activator is used in combination with the inhibitor due to the length of the channel. Different sized circuits would therefore, have different functionality which could be tuned further by the use of additional VOCs with stronger inhibitory or attractive properties. This simple experiment does highlight the hierarchy that exists between chemicals with a known activation (positive chemotaxis) or inhibition (negative chemotaxis) of *P. polycephalum*. Initially when this work was started a relatively strong inhibitor nonanal was selected for use, but his just resulted in complete suppression of the signal regardless of the combination of inputs. Whereas, the relatively weak inhibitor cis-3-hexenyl acetate implemented signal suppression but allowed propagation from the inoculation site, thus differentiating the various inputs and highlighting the subtle balance between inhibitory and activation effects which could be exploited further in future work. It does highlight that even a relatively weak inhibitor as assessed previously by chemotactic assays can still counteract the effects of a strong activator.

The central left hand image of Figure 1 shows the case where no chemical is used as the input (left hand side) and the activator is used as the other input (right hand side). The result is directed signal transfer of *P. polycephalum* towards the activator. The central right hand image shows the opposite case where the activator is on the left of the image and no chemical is on the right. The result is again directed signal propagation towards the activator.

The bottom left hand image in Figure 1 shows the case where the activator is present at both inputs. The result is the propagation of *P. polycephalum* up the input channel, where it splits and moves towards both sources of activator in the output channels. Thus signal splitting is implemented.

The bottom right hand image shows the case where there are no chemical inputs. The result is that a signal propagates up the input channel and then in a random direction towards one output channel.

**Table 1.** Summarises the various inputs and outputs from the simple T-shaped junction, when using *P. polycephalum* as a constant input in the vertical channel and various combinations of chemicals (or absence of chemical input) in the horizontal channels. If the points of chemical input are treated as the outputs for the *P. polycephalum* "signal", then a number of different signal routing operations are implemented. This includes signal suppression, directed signal transfer, random signal generation and signal splitting. **Table 1.** also includes a measure of the consistency/reproducibility of these various operations. Thus signal suppression is reproducibly implemented with *P. polycephalum* failing to reach the output channels in all repeats. The random signal generation is also reproducibly implemented, meaning that when there is no chemical input to the circuit, a single *P. polycephalum* signal is always present at either the right hand or left hand channel. It should be noted that there is no preference for the right hand or left hand channel in the experiments we have undertaken/observed. The success of the circuit in implementing directed signal transfer was measured to be 90%, with 1 in 10 signals either split, failing to propagate to the output or propagating in the wrong direction. The success of the circuit in implementing signal splitting was lower, with 80% of experiments giving the desired result. This just highlights the difficulty of obtaining reproducible results when trying to implement unconventional computing circuits/gates, especially when using a biological entity and diffusive properties of chemicals in combination. However, this error rate compares favourably with other unconventional

approaches such as the construction of gates and circuits in the experimental BZ reaction[39-41]. It obviously doesn't approach the levels that would be required for conventional electronics circuits, although many components in these industries are selected after manufacture based on certain device characteristics/tolerances. Is there any technique to improve the reproducibility of the formed circuits? It is possible that the input strength varies due to the ill-defined amount of culture added to the input channel. Thus methods of standardising the viability and mass of *P. polycephalum* inoculum may impact positively on the results in this and other approaches to forming circuits. It seems like suppression based circuits are easier to implement than activation based. Therefore, design of more complex circuits with higher operational success may be possible by careful assessment of the amount of suppression obtained when blending weak inhibitors with activators and other weak inhibitors etc. Thus it was observed in this work that suppression was reproducible, but a further assessment of whether limited suppression was reproducible and to what level of precision would need to be undertaken. If it were then circuits with various outputs based on the specific level of suppression could be designed.

| Chemical Input | Signal input | Physarum ouput | T-junction Signal output | Pictorial output of circuit | % success | Result |
|---|---|---|---|---|---|---|
| II | -1,-1 | Confined to inoculation site | 0,0 | ■ | 100% | Signal suppression |
| IA | -1, 1 | Confined to input channel | 0,0 | ▌ | 100% | Signal suppression* |
| AI | 1,-1 | Confined to input channel | 0,0 | ▌ | 100% | Signal suppression* |
| IN | -1,0 | Confined to vicinity of inoculation site | 0,0 | ▌ | 100% | Signal suppression* |
| NI | 0,-1 | Confined to vicinity of inoculation site | 0,0 | ■ | 100% | Signal suppression* |
| NN | 0,0 | Ouput at one site only | 1,0 or 0,1 | ↰ ↱ | 100% | Random signal generation/ transfer |
| AN | 1,0 | Ouput at Activator input site only | 1,0 | ↰ | 90% | Directed Signal transfer |
| NA | 0,1 | Output at Activator input site only | 0,1 | ↱ | 90% | Directed signal transfer |
| AA | 1,1 | Output at both activator input sites | 1,1 | ↰↱ | 80% | Signal splitting |

**Table 1.** The various chemical inputs and Physarum polycephalum outputs when using a simple T-shaped junction. Where A=farnesene, I= cis-3-hexenylacetate and N= no chemical input.
*Signal suppression here is based on the size of the circuit rather than being absolute.

**Compound T-shaped Junction**

**Figure 4.** shows selected results from the attempts to implement a more complex routing circuit with a constant *P. polycephalum* input and potentially six chemical inputs. This work showed that it was possible to route a signal via a specific pathway to a predestined output point. For example in Figure 4. We can see in the top right hand image that the signal is routed through the central junction towards the output on the top right of the image only. In this case the circuit has 3 activator inputs (2 in the central region and 1 at the top left hand side where the final signal exits from) and 3 neutral inputs (no chemical). Thus the signal is routed through 3 outputs all of which were sources of the activator chemical farnesene. Thus

the routing circuit is implemented correctly according to the inputs, at the first T-junction we obtain signal splitting at the $2^{nd}$ T-junction we obtain directed signal transfer. Interestingly in this compound circuit after the first junction the presence of no chemical input at the T-junction on the left hand side does not generate a random signal. This is presumably because there is already activator within the circuit so the termination points of the circuit (at least in the short/medium term) are sources of activator at the various inputs. In the original T-shaped junction experiments, the random signal generation arose from a circuit completely free of activator and inhibitor – thus the signal was not suppressed, or directed.

The image on the top left hand side of **Figure 4.** shows the same circuit implemented in another repeat experiment but giving an incomplete output. In this case the *P. polycephalum* correctly implements the central junction (which has two sources of activation) implementing signal splitting, but leaves the confines of the channel and moves towards the activator source external to the channels of the circuit. This was the main error observed for the implementation of this compound junction. This highlights the problem of attempting to confine *P. polycephalum* to this artificial circuit scheme. This is especially true where the channel widths and distance between sources of activator are reduced, meaning that the circuit definition is functionally reduced. This is presumably because Physarum polycephalum wants to minimise the area covered by its tubular network whilst maximising nutrient intake i.e. joining sources of nutrients and/or chemical activators. Where channels are wider there is some incentive to travel on the nutrient substrate to the source of activation, despite this meaning a less direct route. However, it seems apparent that the *P. polycephalum* is able to calculate the relative benefits of moving across a nutrient absent environment to increase speed/reduce distance to source of activation. Alternatively the *P. polycephalum* has no feedback mechanism relating to the environment and is simply moving towards the activator based on the localised air and liquid diffusive gradients, which impact directly on its motor control system. However, if this was the case we might expect the larger but simpler unctions to have the same failure modes. Another common error was the lack of signal splitting at the second T-junction. Therefore, signal splitting was reasonably reproducibly initiated at the central T-junction (within quoted error in **Table 1**) but not at the peripheral junctions. Therefore, although a certain type of behaviour was consistently observed at a simple junction it did not seem to follow that exactly the same behaviour would be reproduced if these simple elements were combined into a more complex circuit. This was true when trying to cascade simple circuit designs in precipitating chemical reactions[42]. Additional work needs to be undertaken in order to understand the subtleties of these interactions. Indeed with better understanding the chance of implementing highly complex circuits based on *P. polycephalum* and its response to chemicals in the environment seem high. Why do we conclude this? There seems to be a definite level of suppression, with various combinations of activators/inhibitors/neutral substances (absence of chemical). There seems to be a hierarchial response at a series of junctions, i.e. the presence of activator within the circuit seems to negate some effects at the additional junctions such as random signal generation and signal splitting. Understanding these mechanisms more clearly could aid in the design of complex signal routing architectures, or alternatively the fine control of the spatial configuration of the plasmodium using chemical sources.

**Control of spatial arrangement of Physarum cultures**

**Figure 5.** shows the results from experiments where *P. polycephalum* is inoculated at a central point within a petri dish and upto four sources of activator (farnesene) are placed at the four points of the compass (N,S,E,W) surrounding the inoculation site. Alternatively, the activator is replaced at 1 or more sites with no chemical i.e. filter paper alone. Therefore,

there are numerous different combinations of spatial inputs. Figure 5 shows that where there are 4 activator inputs the *P. polycephalum* spans and occupies all four sites for an extended time period (middle right hand image). Unlike the T-junction experiments where the culture is stopped from minimising its spatial dimensions via the imposition of channels, in this homogenous media, the *P. polycephalum* approximates a minimal spanning tree of the 4 sites. Also shown are various combinations of two and three inputs showing that *P. polycephalum* preferentially occupies and spans these sites but largely ignores the sites where activator chemical is absent. There is obviously some time dependence to this effect and the *P. Polycephalum* would eventually explore the whole environment once the sources of activator are depleted. The top left hand image shows the case where activator is placed in two locations (S and E). The top right image shows the case where activator is placed in two locations (N and S). The middle left image shows the case where activator is placed in three locations (N and E and W). The bottom right hand image shows the case where activator is placed in 3 locations (S and E and W). There is noticeable segregation of the culture to different localised areas of the dish corresponding to the sources of activator. In many cases the initial exploration phase of the *P. polycephalum* is apparent, growing out uniformly from the inoculation site prior to selecting the final direction of movement. Therefore, an abandoned tubular network can be observed on the scans in addition to the live culture. It is apparent even from this abandoned network that the *P. polycephalum* culture is completely absent over long time periods (>24 hours) from parts of the dish without activator. The case of 1 activator input is not shown, but in this case *P. polycephalum* becomes localised around the source of the chemoattractant chemical. In the case of no inputs the *P. polycephalum* travels to a random site on the periphery of the dish and then usually proceeds to visit each site in turn. Thus there is no localisation over an extended time period as observed when activator is present. Therefore, it is possible for *P. polycephalum* to react to and span multiple activator sources in its environment. The application of chemicals in this manner proves to be a reproducible method of controlling the spatial arrangement of *P. polycephalum* on a homogeneous substrate. It is envisaged that this method can be used for fine, possibly online control of the plasmodiums movement and spatial configuration. This would be useful in the design of computing architectures, geometric calculations (work has been done in this area see for example) and robotic controllers. Taking the last example, in previous work by Tsuda and co-workers they designed various *P. polycephalum* based chips which could be used for the offline and online control of mobile robots[28-30]. The simple but elegant concept behind the chip was to interface the plasmodium directly to an impedance measurement circuit which measured its movement within a confined space. In various light fields experienced by the robot the *P. polycephalum* would alter its spatial configuration moving away from the strongest source of light if it had been differentially applied. This information was fed to the robot control system allowing it to move according to the light field and the programmed operation. It is not difficult to see how this approach could be extended for the measurement of chemical plumes in the environment rather than light levels. We have shown that the *P. polycephalum* changes its spatial arrangement on a macroscopic scale in a reproducible manner according to the position of chemical activators in the environment. It is likely that the same circuit used for the light level work would be able to pick up much smaller microscopic changes in spatial configuration, allowing faster response time. Indeed recent work by our group has shown that changes in the oscillation frequency and amplitude of the *P. polycephalums* naturally generated electrical potential, change specifically according to the applied chemical within its vicinity. This ability to sense chemical plumes and adjust its spatial configuration via localised shifts in oscillation dynamics make the plasmodium a good candidate for the manufacture of robot control chips, provided a better understanding of the mechanisms can be established. In fact due to the induced taxis the plasmodium could be said to already act as an analogue of an amorphous autonomous robot.

**Conclusions**

This work has shown that the routing of *P. polycephalum* through a simple T-shaped junction can be predictably controlled using combinations of chemical inputs. The application of inhibitors with any combination of activator or the absence of chemical resulted in signal suppression (i.e. no output from the current circuit). However, there were levels of suppression dependent on the inputs which may be exploited to design more complex circuits in future work. The application of no chemical input resulted in a random signal generation (i.e. a single output at either of the possible output sites). The application of activator in combination with no chemical resulted in signal routing towards the source of the activator. The application of activator at both inputs resulted in signal splitting and two outputs (i.e. the *P. Polycephalum* moved towards both sources of activator).

The idea was extended in a limited subset of experiments using a compound T-junction which whilst maintaining one *P. polycephalum* input had six possible chemical inputs. This work showed that a *P. polycephalum* signal could be routed reproducibly through a complex junction using chemical inputs. This was particularly the case for 1 or 2 inputs of activator at the central T-junction and 1 activator input at one of the secondary T-junctions. Although even in this case the smaller size of the compound junction meant that the plasmodium was not always confined to the channels but found a more "biologically favourable route". The circuit did not reproducibly implement random signal generation or signal splitting at the secondary T-junctions. This shows that a series of operations from a simple junction can not necessarily be cascaded to multiple junctions, particularly where activator chemicals are already present within the circuit. However, it does highlight that with a better understanding of these effects and size dependent factors it should be possible to predictably and reliably route signals of *P. polycephalum*. This would be useful for designing computing circuits using various activators and inhibitors as inputs.

In additional work it was shown that the spatial configuration of *P. polycephalum* could be reliably altered according to the position of an activator chemical in its environment. In our experiments we tested all possible combinations of activator vs. no chemical. It was found that *P. polycephalum* would adjust its spatial configuration to span four, three and two sources of the activator (in various locations). If just one source was present then *P. polycephalum* would move towards the source and remain localised in the vicinity of the activator for long periods of time. If no sources were present then *P. polycephalum* would adopt a more conventional foraging search approach where it visited all sources in turn, starting with a random but predominantly single location on the outer edge of the dish. This work highlights the possibility of using Physarum polycephalum in a spatial chip on board a mobile robot platform (or in isolation), for applications in chemical plume sensing. Obviously in this case we are utilising macroscopic visible effects but as proven in previous work using light, segregation of the plasmodium, directly interfaced to electrical measurement circuitry could give microscopic spatial data on the positioning and concentration of environmental chemicals.

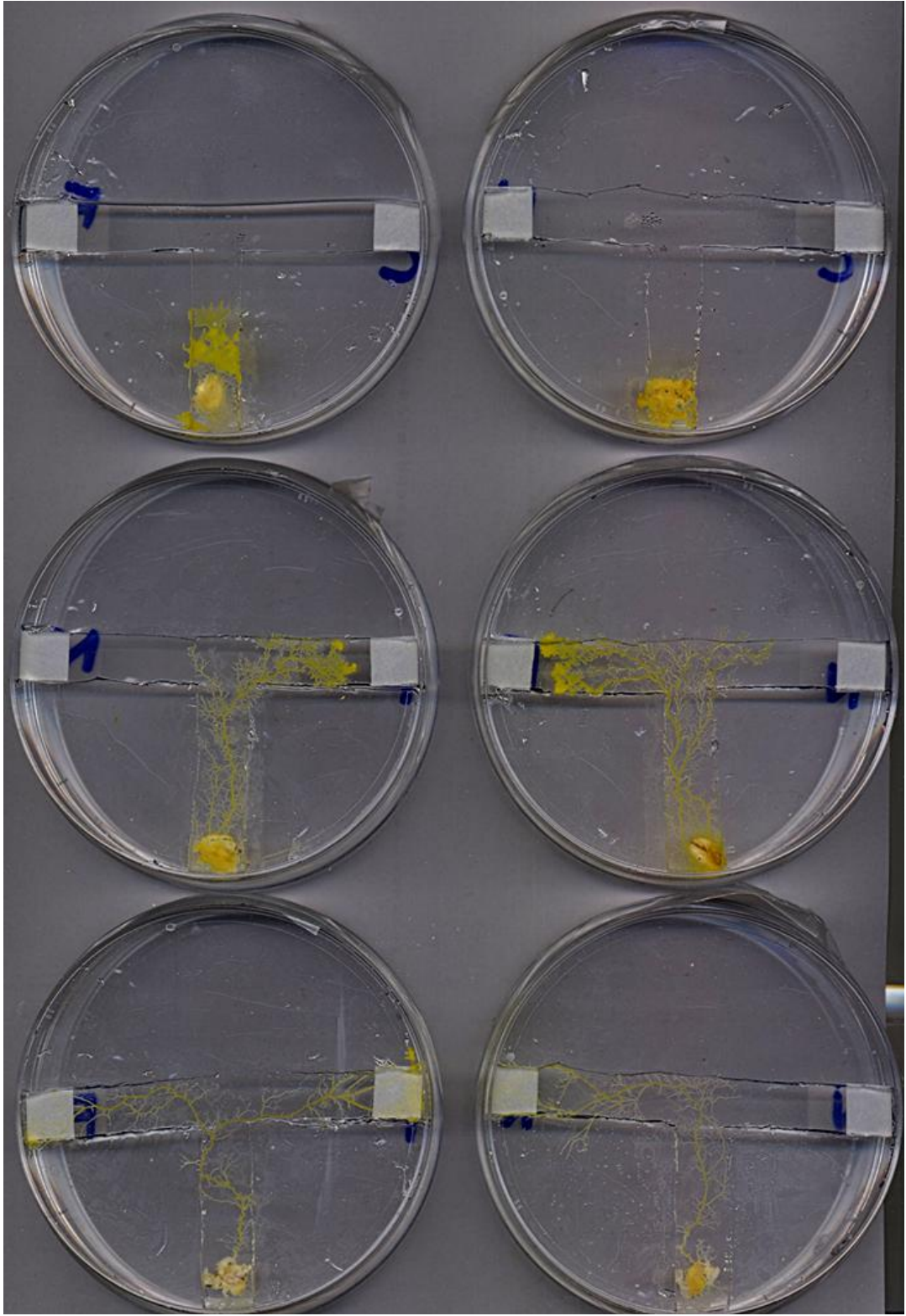

**Figure 3**. Examples of routing of *P. polyc*ephalum through a T shaped junction using simple VOCs. Top left shows the case where farnesene (positive) is added to the left hand output of the junction, whereas cis-3-hexenyl acetate (inhibitor) is added to the right hand output. The result is that *P. polycephalum* although alive is confined to the input channel --- thus the result is signal suppression. The top right image shows the case where no chemical (neutral) is added to the left hand output of the junction, whereas cis-3-hexenyl acetate is added to the right hand output. The result is again signal suppression, albeit stronger. The central left

image shows the case where no chemical is added to the left hand output and farnesene is added to the right hand side. The result is directed signal transfer to the farnesene output. The central right image shows the opposite case where the farnesene input and no chemical input have been reversed. Again the result is directed signal transfer to the farnesene output. The bottom left image shows the case where farnesene has been added to both inputs. The result is signal splitting with the plasmodium directed towards both outputs of the junction. The bottom right image shows the case where no chemical has been added to both inputs. The result is a random signal propagating towards either output.

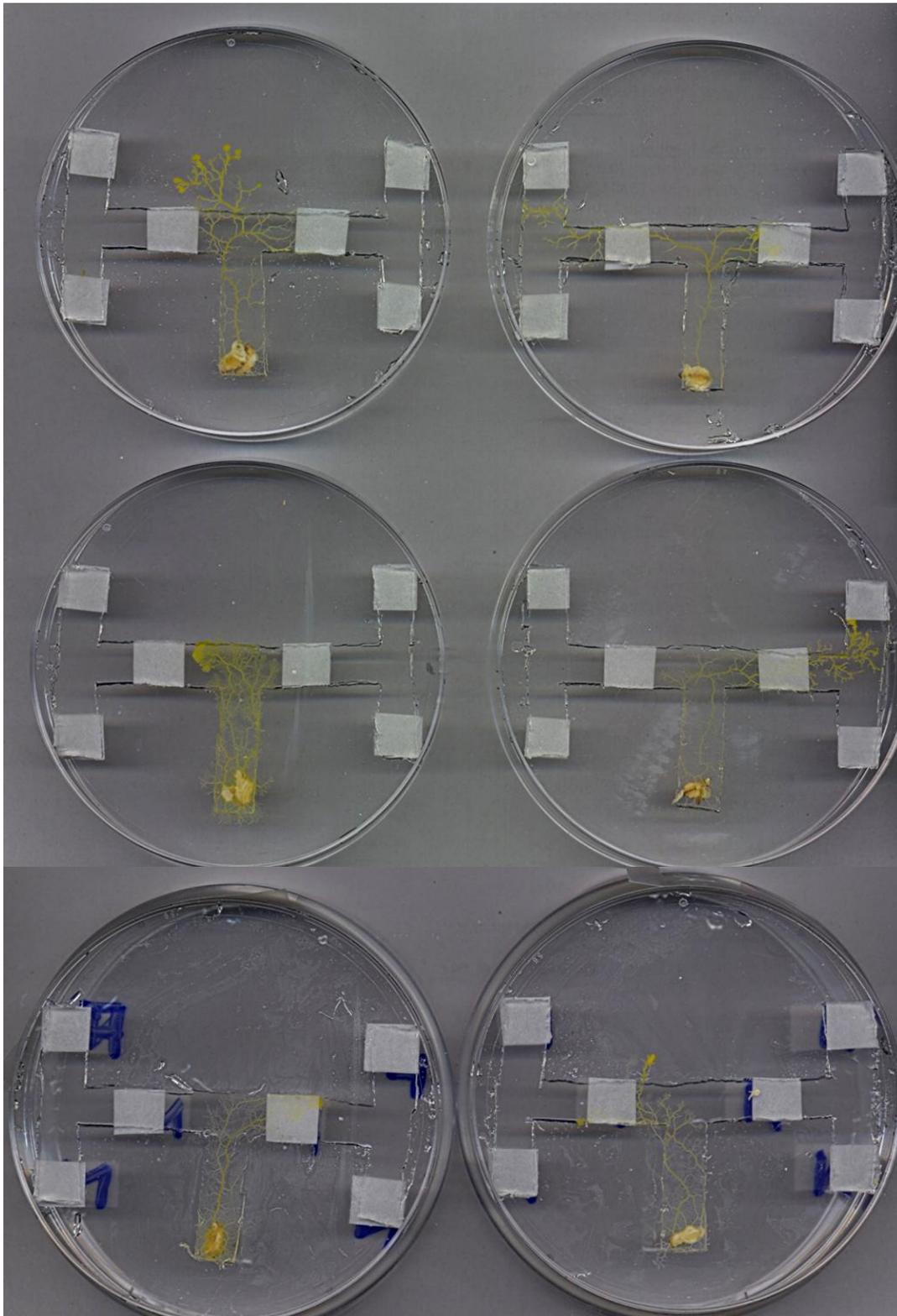

**Figure 4**. Selected results from the implementation of a compound T-shaped junction with 6 possible sites of chemical inputs. The top left hand side image shows the case where farnesene (Activator) is present at both central inputs and also in the input at the top left hand side. This hsows that Physarum polycephalum appears to take a cross country route towards the correct output configuration. The top right hand side image shows a circuit with the same inputs which is correctly implemented via the channels. The central left hand side image shows a circuit where Farnesene is present at both central inputs, but no chemical input is present at the other inputs. The result is that the Physarum signal remains confined in the central region for an extended period of time >24 hours. The central right hand side image shows a circuit where farnesene is present at only the right hand side central input and also the top right hand side input. The signal is successfully transferred through the circuit The bottom left hand side image shows the case where farnesene is only present at the central right hand side input, whereas the bottom right hand side image shows the same circuit with farnesene at the central left hand side input. In both cases it shows that there is directed transfer through the first part of the circuit then segregation in the central portion for an extended time period.

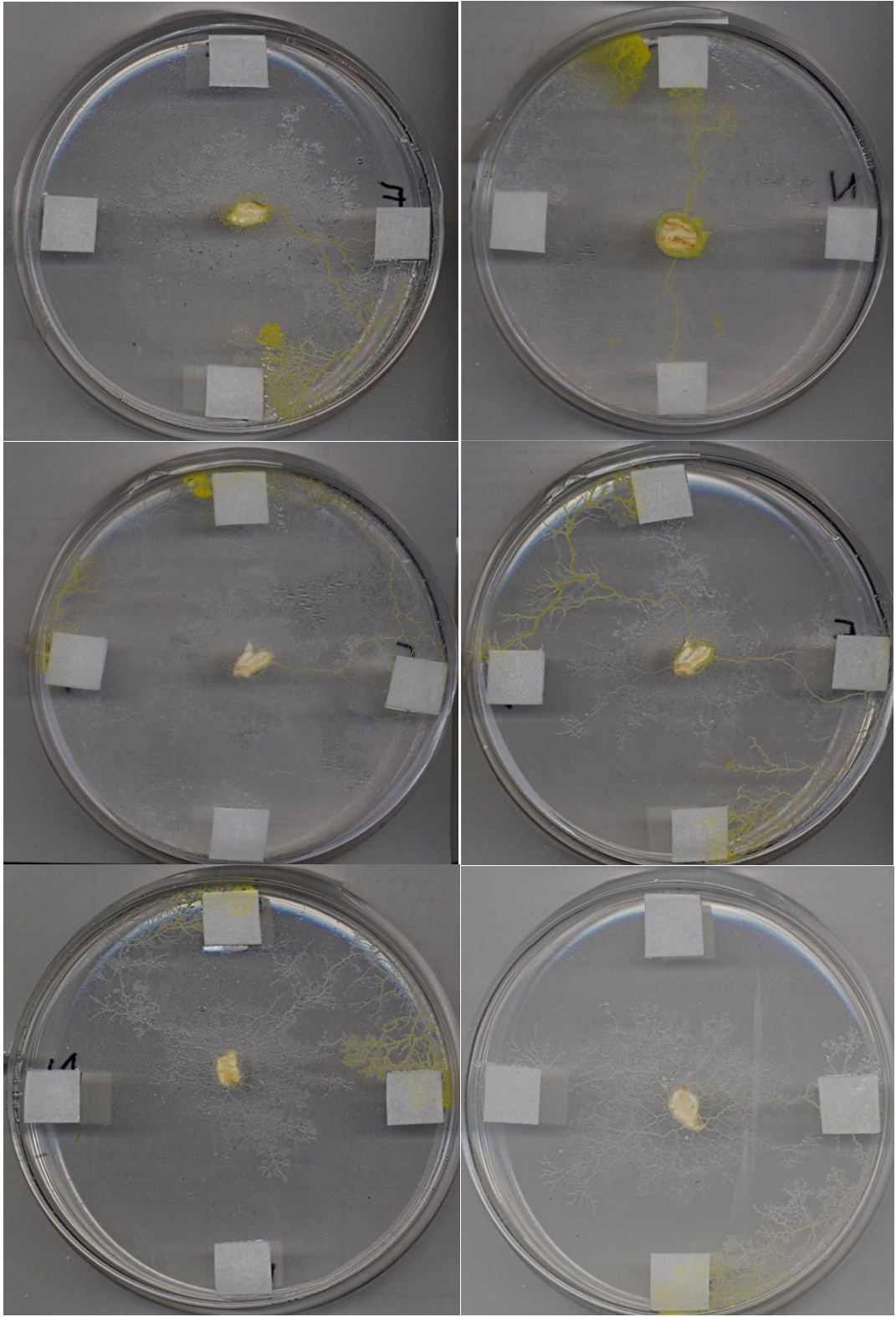

**Figure 5**. Showing the control of the spatial configuration of Physarum using presence or absence of an activator chemical at four specific points surrounding the inoculation site. The top left hand image shows the case where activator is placed in two locations (S and E). The top right image shows the case where activator is placed in two locations (N and S). The middle left image shows the case where activator is placed in three locations (N and E and W). The middle right hand image shows the case where activator is placed in four locations.

The bottom left hand image shows the case where activator is placed in two locations (N and E). The bottom right hand image shows the case where activator is absent placed in 3 locations (S and E and W).